\documentclass[final,5p,times,twocolumn,sort&compress]{elsarticle}
\usepackage{booktabs}
\usepackage{tabularx} 
\usepackage{amsmath}
\usepackage{enumitem}
\setlist{nosep}
\usepackage[colorlinks, citecolor=blue, urlcolor=blue, linkcolor=blue, bookmarksdepth=3]{hyperref}
\usepackage{etoolbox}
\apptocmd{\thebibliography}{\raggedright}{}{}
\usepackage{graphicx}
\usepackage{microtype}
\usepackage{xcolor}
\usepackage{courier}
\usepackage{listings}
\lstdefinelanguage{G4macro}{
  basicstyle=\ttfamily\footnotesize,
  morecomment=[l]{\#},
  keywordstyle=[1]\color{cyan!70!black},
  keywordstyle=[2]\color{teal},
  keywordstyle=[3]\color{green!50!black},
  keywordstyle=[4]\color{magenta},
  alsoletter={/, >},
  morekeywords=[1]{
    /control/, /units/, /particle/, /tracking/, /geometry/, /process/, /event/, /cuts/, /run/, /random/, /control/cout/, /vis/, /analysis/, /material/, /physics_lists/, /score/, /globalField/, /param/,
    /control/verbose, /gui/clearMenu, /gui/addMenu, /gui/addButton, /geometry/test/run, /score/create/realWorldLogVol, /score/quantity/doseDeposit, /score/quantity/energyDeposit, /score/close, /score/dumpQuantityToFile, /score/drawProjection, /globalField/setValue, /gui/system, /control/execute, /gps/particle, /gps/energy, /gps/direction, /gps/pos/centre, /run/initialize, /run/numberOfThreads, /run/beamOn, /process/em/verbose, /process/had/verbose, /globalField/verbose, /run/verbose, /control/verbose, /tracking/verbose, /vis/open, /vis/drawVolume, /vis/scene/add/axes, /vis/scene/add/gps, /vis/scene/add/scale, /vis/scene/add/magneticField, /vis/scene/add/trajectories, /vis/viewer/set/background, /vis/viewer/set/lightsVector, /vis/viewer/set/upVector, /vis/viewer/set/viewpointVector, /vis/viewer/set/auxiliaryEdge, /vis/scene/endOfEventAction, /vis/viewer/zoom, /vis/ogl/export, /control/manual, /control/createHTML, /analysis/list, /material/nist/listMaterials, /particle/list, /process/list, /units/list, /score/list, /score/colorMap/listScoreColorMaps,
    /control/doifBatch, /control/echo, /control/doifInteractive, /control/getEnv, /control/listAlias, /run/printProgress, /control/cout/useBuffer, /run/verbose, /vis/list, /vis/scene/add/trajectories, /vis/viewer/set/background, /vis/scene/endOfEventAction, /vis/viewer/zoom, /vis/ogl/export, /process/em/verbose, /process/had/verbose, /globalField/verbose, /analysis/list, /material/nist/listMaterials, /particle/list, /process/list, /units/list, /score/list, /score/colorMap/listScoreColorMaps, /score/create/realWorldLogVol, /score/quantity/doseDeposit, /score/quantity/energyDeposit, /score/close, /score/dumpQuantityToFile, /score/drawProjection
  },
  morekeywords=[2]{cm, m, mm, keV, MeV, GeV, tesla},
  morekeywords=[3]{PreInit>, Idle>},
  morekeywords=[4]{G4UI_USE_TCSH, G4RUN_MANAGER_TYPE, PHYSLIST, CMAKE_EXPORT_COMPILE_COMMANDS, GEANT4_DATA_DIR, PATH, ls, cd}
}

\lstdefinelanguage{G4C++}{
  language=C++,
  morecomment=[l][\color{green!50!black}]{\#include},
  morekeywords=[2]{G4VUserDetectorConstruction, G4GlobalMagFieldMessenger, G4tgbVolumeMgr, G4AutoDelete, G4VPhysicalVolume, G4VUserPrimaryGeneratorAction, G4GeneralParticleSource, G4Event, G4VUserActionInitialization, G4TScoreNtupleWriter, G4RunManagerFactory, G4PhysListFactory, G4AnalysisManager, G4ScoringManager, G4VisExecutive, G4UIExecutive, G4UImanager, G4RunManager, G4String},
}

\lstdefinelanguage{Tg}{
  basicstyle=\ttfamily\footnotesize,
  morecomment=[l]{\#},
  alsoletter={:,-},
  morekeywords=[1]{:volu, :rotm, :place, :vis, :color, :prop, :property, :surf, :photon_energies, :const, :type, :model, :finish, :SigmaAlpha, \#include},
  keywordstyle=[1]\color{purple},
  morekeywords=[2]{G4_AIR, G4_WATER, G4_A-150_TISSUE, G4_BONE_COMPACT_ICRU, OFF},
  keywordstyle=[2]\color{blue},
  morekeywords=[3]{BOX, CONS, TRD, TUBS, ORB, SPHERE, PARA, TRAP, TORUS},
  keywordstyle=[3]\color{teal}
}

\lstdefinelanguage{Root}{
  language=C++,
  alsoletter={.},
  morekeywords=[1]{TFile, TTree},
  keywordstyle=[1]\color{blue},
  morekeywords=[2]{Show, Draw, .ls},
  keywordstyle=[2]\color{magenta},
  moredelim=[s][\color{green!50!black}]{root[}{]}
}
\makeatletter
\def\verbatim@font{\ttfamily\small}
\makeatother

\journal{Computer Physics Communications}

\begin{document}

\begin{frontmatter}
\title{MinGLE: A Minimalist, Configurable, and Pedagogical Geant4 Application Template}
\author{Jing Liu}
\ead{jing.liu@usd.edu}
\affiliation{organization={Department of Physics, University of South Dakota},
            addressline={414 E. Clark St.}, 
            city={Vermillion},
            postcode={57069},
            state={SD},
            country={USA}}

\begin{abstract}
The Geant4 toolkit is the standard for simulating particle transport through matter, yet its steep learning curve often hinders novice developers. This paper introduces MinGLE (Mini Geant4 Learning Example), a minimalist application template designed to bridge the gap between being a Geant4 end user and a developer. By leveraging modern factory classes, text geometry syntax, and the general particle source, MinGLE provides a complete, functional simulation kernel in under 70 lines of C++ code. Its pedagogical innovation lies in an incremental development roadmap formalized through ten testable Git branches and tags. Each stage introduces a core Geant4 component, from user interfaces and physics to scoring and magnetic fields, ensuring that every line of code is motivated by a specific requirement. MinGLE serves as both a powerful, universal starting point for new projects and a step-by-step educational resource, transforming passive documentation into an active, hands-on development experience.
\end{abstract}

\begin{keyword}
Geant4, Monte Carlo simulation, Application template, Text geometry, Pedagogy, Git, GitHub, Minimalist design
\end{keyword}

\end{frontmatter}

\section{Introduction}
Geant4 \cite{g403, g406, g416} is the standard software for simulating the passage of particles through matter, serving as a critical resource for detector design and data analysis across various scientific disciplines. Geant4 positions itself explicitly as a toolkit, obligating users to act as Geant4 \emph{application developers} instead of \emph{end users} in that they must understand and correctly assemble several core Geant4 components (run management, physics, geometry, etc.) to create a functional executable. Consequently, the initial learning curve for Geant4 is steep.

Official Geant4 examples are designed primarily to demonstrate assembly of various Geant4 components but are frequently tied to specific detector models, requiring significant effort to deconstruct into a universal template.

To lower this entry barrier, projects like the Geant4 Example Application with Rich features yet Small footprint (GEARS) \cite{gears} have sought to shield users from C++ development altogether. By enabling the comprehensive configuration of a simulation through external text files, GEARS allows beginners to operate as \emph{end users} of a pre-built application rather than as developers tasked with assembling a toolkit from scratch.

The Geant4 Collaboration has also recognized that the traditional reliance on complex, low-level component assembly is error-prone and requires users to master excessive specific knowledge. Modern components, such as factory classes (e.g., for physics lists and run managers), provide a cleaner, abstracted interface, effectively performing some of this assembly internally. However, these classes are still components of the toolkit. They need to be explicitly called and assembled in C++ code, and this does not convert Geant4 into an \emph{end-user-proof} application like GEARS.

This paper introduces MinGLE (Mini Geant4 Learning Example) \cite{mingle}, an application template designed to bridge the gap between being a Geant4 \emph{end user} and a Geant4 \emph{application developer}. MinGLE is similar to GEARS in that all essential configurations are done at runtime through macro and text files. However, unlike GEARS, MinGLE contains no Geant4 extension code; it relies exclusively on standard Geant4 components and interfaces.

MinGLE serves two primary purposes:
\begin{itemize}
    \item To provide a clean, accessible starting point for developers who prefer a minimalist, standard Geant4 application core.
    \item To act as a transitional learning tool for \emph{end users} familiar with runtime configuration (e.g., from using GEARS) who now wish to understand and modify the underlying C++ source code.
\end{itemize}

MinGLE provides a complete, contemporary Geant4 application template, \emph{mingle.cc}, with fewer than 70 lines of C++ code shown entirely in Listing \ref{l:mingle}. This minimalism is achieved by leveraging modern Geant4 factory classes and external configuration files, ensuring the program is highly configurable, universal, and easily studied.

It is worth noting that even though MinGLE is primarily provided as a template for extension, it already includes all Geant4 essential components to perform a variety of simulation and analysis tasks as it is.

\begin{figure*}[htbp]\centering
\begin{lstlisting}[language=G4C++, caption={Universal, single-file Geant4 application template, \emph{mingle.cc}, in its entirety.}, label={l:mingle}, numbers=left, frame=leftline, xleftmargin=5em]
#include <G4VUserDetectorConstruction.hh>
#include <G4GlobalMagFieldMessenger.hh>
#include <G4tgbVolumeMgr.hh>
#include <G4AutoDelete.hh>
class Detector : public G4VUserDetectorConstruction
{
  public:
    G4VPhysicalVolume* Construct() {
      G4tgbVolumeMgr::GetInstance()->AddTextFile("detector.tg");
      return G4tgbVolumeMgr::GetInstance()->ReadAndConstructDetector();
    } ///< load detector definition from a text file "detector.tg"
    void ConstructSDandField() {
      G4AutoDelete::Register(new G4GlobalMagFieldMessenger());
    } ///< enable /globalField/ to set uniform B-field
};

#include <G4VUserPrimaryGeneratorAction.hh>
#include <G4GeneralParticleSource.hh>
class Generator : public G4VUserPrimaryGeneratorAction
{
  private:
    G4GeneralParticleSource* fGPS;
  public:
    Generator() : G4VUserPrimaryGeneratorAction() { fGPS = new G4GeneralParticleSource; }
    ~Generator() { delete fGPS; }
    void GeneratePrimaries(G4Event *evt) { fGPS->GeneratePrimaryVertex(evt); }
};

#include <G4VUserActionInitialization.hh>
class Action : public G4VUserActionInitialization
{
  public:
    void Build() const { SetUserAction(new Generator); }
};

#include <G4TScoreNtupleWriter.hh>
#include <G4RunManagerFactory.hh>
#include <G4PhysListFactory.hh>
#include <G4AnalysisManager.hh>
#include <G4ScoringManager.hh>
#include <G4VisExecutive.hh>
#include <G4UIExecutive.hh>
#include <G4UImanager.hh>
int main(int argc, char** argv)
{
  auto run = G4RunManagerFactory::CreateRunManager();
  G4ScoringManager::GetScoringManager();  // activate command-based scorer
  G4TScoreNtupleWriter<G4AnalysisManager> writer; // enable ntuple recording
  if (run->GetRunManagerType() != G4RunManager::sequentialRM)
    writer.SetNtupleMerging(true); // merge ntuples created in multi-threads
  // load default physics list, or the one specified by $PHYSLIST
  G4PhysListFactory f; run->SetUserInitialization(f.ReferencePhysList());
  run->SetUserInitialization(new Detector); // load detector definition
  run->SetUserInitialization(new Action); // load particle generator

  G4UIExecutive* ui = nullptr; // assume batch mode by default
  if (argc==1) ui = new G4UIExecutive(argc, argv); // interactive mode
  // enable visialization between UI creation and session start
  auto vis = new G4VisExecutive("quiet"); vis->Initialize();
  if (ui) { // start an interactive UI session
    ui->SessionStart(); delete ui; // clear memory at the end of UI session
  } else { // run a macro file in batch mode
    G4String cmd = "/control/execute ", macroFile = argv[1];
    G4UImanager::GetUIpointer()->ApplyCommand(cmd + macroFile);
  }
  delete vis;
  delete run;
}
\end{lstlisting}
\end{figure*}

\begin{table*}[t]\centering\small
\caption{Guided Geant4 application development roadmap labeled with Git branches and tags.}\label{t:map}
\begin{tabularx}{\textwidth}{@{}clll X@{}}
        \toprule
        \textbf{Stage} & \textbf{Milestone} & \textbf{Branch} & \textbf{Tag} & \textbf{Focus} \\
        \midrule
        0 & user interface        & ui       & v0 & interact with Geant4 via TUI/GUI and navigate the macro command tree \\
        1 & batch execution       & batch    & v1 & transition from interactive exploration to automated batch-mode execution \\
        2 & run manager           & run      & v2 & initialize the simulation kernel and understand the lifecycle of a simulation \\
        3 & physics list          & physics  & v3 & use G4PhysListFactory to select validated, high-level physics processes \\
        4 & detector definition   & detector & v4 & define detector geometry and materials using simple syntax in text files \\
        5 & visualization         & vis      & v5 & implement G4VisExecutive to render 3D geometry and trajectories \\
        6 & particle generation   & gps      & v6 & control particle generation dynamically with the General Particle Source \\
        7 & histogram generation  & scorer   & v7 & use built-in command-line mesh scorers to generate statistical distributions \\
        8 & ntuple generation     & ntuple   & v8 & use G4AnalysisManager to save event-by-event data in ntuples \\
        9 & magnetic field        & field    & v9 & add uniform B-field to simulate particle deflection \\
        \bottomrule
    \end{tabularx}
\end{table*}

\section{Code Management}
MinGLE is hosted on GitHub \cite{github} to leverage the modern software development ecosystem, utilizing Git \cite{git} and GitHub not only for version control, code hosting and distribution, but also as pedagogical tools for novice application developers.

\subsection{Git as a Pedagogical Tool}
There are two primary ways to teach a complex software framework like Geant4. The first is to disassemble a finished, feature-complete application to see how it works. However, for a beginner, this reverse engineering often feels like trying to learn how an engine works by looking at a fully assembled car. The second way, and the one MinGLE adopts, is to evolve the code step-by-step. By building the application one component at a time, one can see the specific necessity of every line of code as it is added. This bottom-up approach ensures that one is never overwhelmed by boilerplate, as each stage introduces exactly one new concept.

To facilitate this step-by-step journey, MinGLE utilizes two core features of the Git version control system: branches \cite{br} and tags \cite{tag}, to form a guided development roadmap as shown in Table \ref{t:map}. Each branch in MinGLE contains the addition of a distinct Geant4 component or feature, fully functional and testable, with the tip of it tagged as v$n$, representing a milestone of the completion of that stage of development. This structure serves the following pedagogic purposes:
\begin{description}
\item[Isolation of Concepts] By switching to a specific branch, one can see only the code required for that milestone, free from the ``noise'' of yet-to-be-introduced concepts.
\item[Active Experimentation] Unlike static code snippets, a branch is a live workspace. One can modify the code and commit one's own changes to see how they affect the simulation.
\item[Stable Checkpoints] Tags provide a ``correct answer''. If one's experiments in a branch lead to errors, one can always compare their work against the corresponding tag to find the solution. 
\end{description}

\subsubsection{Git Tag for Static Inspection}
Due to the version control nature of Git, after obtaining a local copy of the repository using the following command
\begin{lstlisting}
git clone https://github.com/jintonic/mingle
\end{lstlisting}
one can always use
\begin{lstlisting}
git log
\end{lstlisting}
to inspect the entire development history of the repository. Each snapshot in the history can be retrieved via a commit ID made up of a series of random characters, for example,
\begin{lstlisting}
git checkout b196f6
\end{lstlisting}
However, this raw, chronological history is non-linear, messy, includes numerous commits for debugging, refactoring, and sometimes leading to dead-ends. Browsing such a history is confusing and counterproductive for a novice user. 

A beginner is adviced to use the \verb|tag| command instead to list the carefully curated development milestones with meaningful names:
\begin{lstlisting}
git tag
v0
v1
...
\end{lstlisting}
the \verb|show| command to inspect a specific milestone:
\begin{lstlisting}
git show v0:mingle.cc
\end{lstlisting}
and the \verb|diff| command to examine the incremental changes between two milestones:
\begin{lstlisting}
git diff v0 v1 mingle.cc
\end{lstlisting}

\subsubsection{Git Branch for Dynamic Exploration}
For advanturous users who are not afraid of changing the code, they can obtain a list of all branches using
\begin{lstlisting}
git branch -a
* main
  remotes/origin/HEAD -> origin/main
  remotes/origin/batch
  ...
  remotes/origin/main
  ...
\end{lstlisting}
where the \verb|*| symbol indicates the current local branch, and the \verb|-a| option lists both local and remote branches.

A freshly cloned repository only has one local branch, the \verb|main| branch. To switch the working directory to another branch, such as \verb|batch|, one can use
\begin{lstlisting}
git switch batch
\end{lstlisting}
which will pull the branch from the remote repository and switch the working directory to it.

If users have modified the code while exploring and try to switch back, Git may prevent the switch to protect their work. In such cases, they can either discard the change to return to a clean state using
\begin{lstlisting}
git restore .
git switch main
\end{lstlisting}
or save the change before switching using
\begin{lstlisting}
git stash
git switch main
\end{lstlisting}
and retrieve the change back later using
\begin{lstlisting}
git stash pop
\end{lstlisting}

\subsection{GitHub as a Pedagogical Tool}
Learning the fundamentals of version control through \verb|git| is a key first step for any aspiring developer. However, it should not be mandatory for a beginner to master \verb|git| to learn Geant4. Hosting the MinGLE git repository on GitHub allows users to achieve the same goal through GitHub's web interface, without the need to master \verb|git| commands.

The MinGLE repository on GitHub, \url{https://github.com/jintonic/mingle}, lists all files in the repository, and users can click on any file to view its content. It also renders the README file in Markdown~\cite{md} format into a well-formatted easy-to-read web page. Markdown is also a plain text format easily readable in any simple text editor on a local computer. It is chosen to ensure optimal presentation both online and offline.

Contents of README files are branch-specific. This transition from a monolithic documentation style significantly enhances the pedagogical value of the repository. By designating the \verb|main| branch README as a high-level ``syllabus'', the project provides a clear executive summary of the MinGLE philosophy and the central roadmap for navigation. Meanwhile, the individual branch READMEs function as specialized ``chapters'', offering deep dives into specific concepts exactly when the user requires that information. Similarly, an example Geant4 macro file \emph{run.mac} is also branch-specific, and users can click on it to view the content of the file. This ``just-in-time'' learning approach creates a logical and cumulative learning path that reduces cognitive load by isolating complexity within its relevant development stage.

The static browsing of tags and branches of the repository can be done using the branch and tag selectors on top of the file list. They can also be directly accessed using the URL:
\url{https://github.com/jintonic/mingle/tree/v0}
where \verb|v0| can be replaced by any tag or branch name.

Each stage of the code can be downloaded as a zip or tar.gz file for inspection or testing from \url{https://github.com/jintonic/mingle/tags}.

GitHub also provides a practical introduction to distributed computing. It forces users to maintain synchronized backups of their work in at least two independent locations: the GitHub server and their local computer, thereby teaching robust data management practices from day one.

Finally, for those intending to use MinGLE as a clean template for their own new development, GitHub allows them to fork the MinGLE repository to create their own independent repository also on GitHub.

\section{Compilation and Deployment}\label{s:cd}

\begin{figure*}[ht]\centering
\begin{lstlisting}[language=CMake, caption={\emph{CMakeLists.txt} ensuring a Release type of executable and its installation to Geant4's bin/ directory.}, label={l:cmake}, numbers=left, xleftmargin=3em, frame=leftline]
cmake_minimum_required(VERSION 3.16) # CMake quits without it

project(mingle) # project name

# http://geant4-userdoc.web.cern.ch/geant4-userdoc/UsersGuides/InstallationGuide/html/buildtools.html
find_package(Geant4 REQUIRED ui_all vis_all) # search for Geant4Config.cmake
include(${Geant4_USE_FILE}) # Linux cannot find Geant4 header files without it

if(WIN32) # set config type for multi-configuration generators, e.g. Visual Studio, Xcode
  set(CMAKE_CONFIGURATION_TYPES Release CACHE STRING "Release;Debug;..." FORCE)
else() # set default build type for single-configuration generators, e.g. Makefile
  if(NOT CMAKE_BUILD_TYPE) # if not defined by the user
    set(CMAKE_BUILD_TYPE Release CACHE STRING "Release;Debug;..." FORCE)
  endif()
endif() # executable of Release type is leaner & faster than that of Debug type

if(CMAKE_INSTALL_PREFIX_INITIALIZED_TO_DEFAULT) # if not defined by the user
  set(CMAKE_INSTALL_PREFIX "${Geant4_DIR}/../../.." CACHE PATH "Installation prefix" FORCE)
endif() # set default installation prefix

if(APPLE) # https://gitlab.kitware.com/cmake/community/-/wikis/doc/cmake/RPATH-handling
  set(CMAKE_INSTALL_RPATH "${Geant4_DIR}/../..")
endif()  # set RPATH in executable on Mac (must be before add_executable(...))

add_executable(${PROJECT_NAME} ${PROJECT_NAME}.cc)
target_link_libraries(${PROJECT_NAME} ${Geant4_LIBRARIES})
install(TARGETS ${PROJECT_NAME}) # enable cmake --install
\end{lstlisting}
\end{figure*}

MinGLE is designed to be runnable across major operating systems (OS) and environments to ensure maximum accessibility for beginners.

The application is shipped with a standard \emph{CMakeLists.txt} file as shown in Listing \ref{l:cmake}, which allows developers to easily compile the source code into an executable on all three major platforms: Windows, macOS, and Linux using CMake \cite{cmake}.

As the build process creates a lot of temporary files, it is recommended to use a separate \verb|build| directory to keep the source directory clean. CMake provides a simple way to do this using the \verb|-B| option, which specifies the directory where \textbf{B}inary executables can go: 
\begin{lstlisting}
cmake -B build 
\end{lstlisting}
Note that the command is case-sensitive. \verb|-B| cannot be replaced by \verb|-b|.

To build (compile) the application, run the following command universally on Linux, macOS, and Windows:
\begin{lstlisting}
cmake --build build --config Release
\end{lstlisting}
On Linux or macOS, a shorter command can be used instead:
\begin{lstlisting}
make -C build
\end{lstlisting}
On Windows, the same commands can be issued in a Git Bash terminal \cite{bash} that comes with a standard installation of Git for Windows. These operations can also be done through graphic user interfaces (GUI) of CMake and Visual Studio \cite{vs} for users who are intimidated by the command line. Detailed instructions on how to do this can be found in the author's YouTube channel \cite{physino}.

Option \verb|--config Release| is used to specify the configuration type for multi-configuration generators, e.g. Visual Studio, Xcode, which can be \verb|Release|, \verb|Debug|, \verb|MinSizeRel|, \verb|RelWithDebInfo|, or any custom type. This is not necessary for single-configuration generators, e.g. Makefile used on Linux or macOS. Line 9 to 15 of the \emph{CMakeLists.txt} file shown in Listing \ref{l:cmake} are used to specify \verb|Release| as the default build and configuration type for users who do not know the difference between \verb|Release| and \verb|Debug| and would likely accept this default setting in Visual Studio on Windows. Generally speaking, an executable of \verb|Release| type is leaner and faster than that of \verb|Debug| type. Even thought this is not obvious for a program as tiny as MinGLE, the benefit of which will be more apparent for a larger application developed on top of MinGLE.

This default setting in \emph{CMakeLists.txt}, however, is ignored by \verb|cmake --build| in the Git Bash terminal on Windows. Ignoring \verb|--config Release| in this niche case may lead to unexpected behavior. Windows terminal users are adviced to always specify it. 

To install the application, run the following command universally on Linux, macOS, and Windows:
\begin{lstlisting}
cmake --install build
\end{lstlisting}
or build the \verb|INSTALL| target through the Visual Studio GUI.

Line 17 to 19 in \emph{CMakeLists.txt} shown in Listing \ref{l:cmake} instructs this command to copy the generated mingle executable into the \verb|bin| directory of the local Geant4 installation detected by CMake in the build process. Since a successful Geant4 installation already requires this directory to be in the user's PATH environment variable, this setting makes mingle globally available from any location without additional environment configuration. A user can hence simply type
\begin{lstlisting}
mingle
\end{lstlisting}
in a terminal or even the Windows file browser's address bar to launch the application.

While designed for simplicity, MinGLE also supports advanced development workflows by leveraging the JSON Compilation Database standard \cite{json}. By setting the environment variable
\begin{lstlisting}
CMAKE_EXPORT_COMPILE_COMMANDS=ON
\end{lstlisting}
during the configuration step, CMake automatically generates a \emph{compile\_commands.json} file in the build directory \cite{cmake}.

This file acts as a bridge between the build system and modern Integrated Development Environments (IDEs) or text editors, such as Visual Studio Code \cite{vscode}, Neovim \cite{neovim}, or CLion \cite{clion}, etc. These editors use the compilation database to resolve Geant4-specific header paths and compiler flags automatically. For a developer, this enables IntelliSense features, including real-time syntax highlighting, accurate code navigation (e.g., ``Go to Definition''), and auto-completion for complex Geant4 classes, without requiring manual configuration of editor-specific include paths. By keeping this configuration in the user's environment (e.g., \emph{\textasciitilde/.zshenv} or \emph{\textasciitilde/.bashrc}) rather than hard-coding it into \emph{CMakeLists.txt}, the project maintains a minimalist core while providing a modern environment for serious application development.

To address the common challenge where new users may struggle with local compilation dependencies (e.g., Geant4 installation, system libraries), a pre-compiled executable is provided together with the entire Geant4 libraries via a Docker image \cite{docker} hosted on Docker Hub \cite{g4img}. This allows users to experience the full functionality of the application immediately, even when they fail to compile it by themselves. Using the Docker container also provides an out-of-the-box experience where the user is shielded from OS-specific compilation issues, enabling them to focus on learning the Geant4 application's development and runtime configuration. 

\section{Development Roadmap}
The development roadmap is organized around ten key stages, each marked with a descriptive Git branch and tag name shown in Table \ref{t:map}. The subsequent subsections walk through this development history step by step, demonstrating the minimal code increment required for each stage.

\subsection{Geant4 User Interface}
The first stage of \emph{mingle.cc} only contains 6 lines of C++ code, centering on the \verb|G4UIExecutive| class \cite{G4UIExecutive}:
\begin{lstlisting}[language=G4C++, numbers=left, xleftmargin=3em, frame=leftline]
#include <G4UIExecutive.hh>
int main(int argc, char** argv)
{
  G4UIExecutive ui(argc, argv);
  ui.SessionStart();
}
\end{lstlisting}
MinGLE at this stage does not function as a simulation program due to the lack of detector and physics components, etc. But it is perfectly executable and provides several immediate, hands-on learning opportunities regarding Geant4 interactive sessions and macro command structure.

Geant4 macro commands are text-based instructions that allow users to control almost every aspect of the simulation at runtime without writing or compiling additional C++ code. This mechanism is one of the primary ways MinGLE achieves high configurability while maintaining a minimalist C++ core.

These commands are organized in a hierarchical tree structure, similar to a Unix filesystem, where related functionalities are grouped into directories such as \verb|/control/|, \verb|/units/|, \verb|/gui/|, etc. At this stage these are the only three directories (or command sets) available. They expand as the program evolves.

\begin{figure*}[t]
\includegraphics[width=\textwidth]{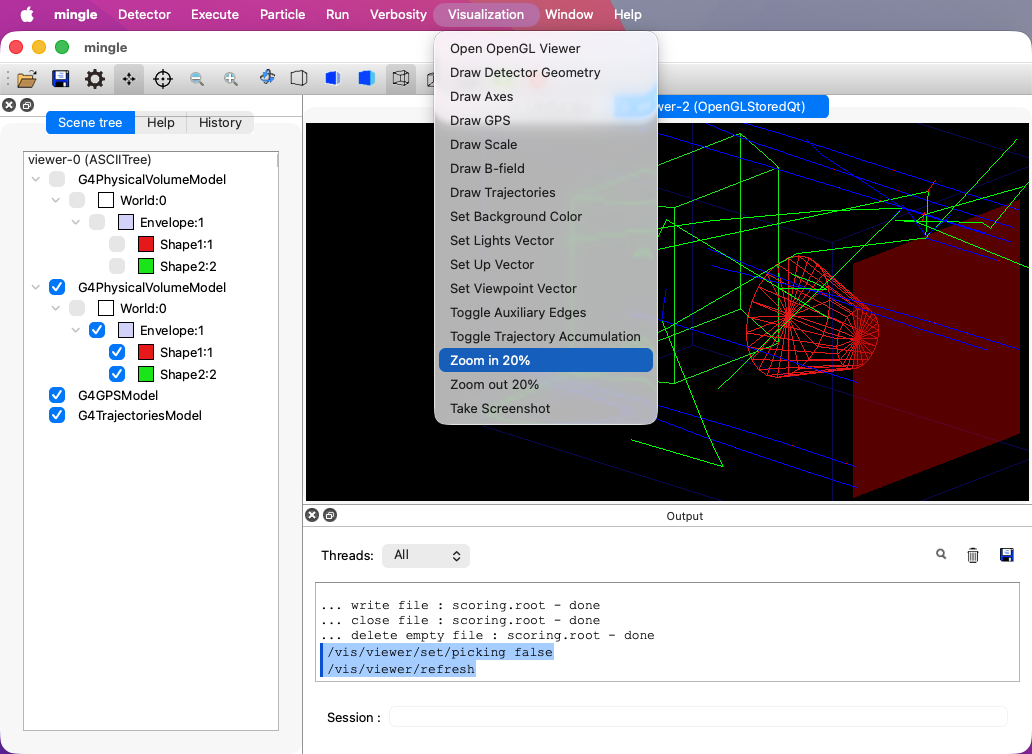}
\caption{Screenshot of the Geant4 GUI with a customized menu, visualized detector geometry, and accumulated particle trajectories.}\label{f:gui}
\end{figure*}

By default, running $\verb|mingle|$ will open a Geant4 GUI window (If that is enabled in the user's local Geant4 installation), providing a visual way to interact with the environment. A screenshot of the GUI in a Mac is shown in Figure \ref{f:gui}.

Users can easily switch to a text-based user interface (TUI) by setting the environment variable \verb|G4UI_USE_TCSH| before execution. The resulting session allows users to interactively navigate the hierarchical Geant4 macro command tree using filesystem-like \verb|ls| and \verb|cd| commands as shown below:
\begin{lstlisting}[basicstyle=\ttfamily\scriptsize, language=G4macro]
G4UI_USE_TCSH=1 mingle
Available UI session types: [ Qt, tcsh, csh ]
PreInit> ls
Command directory path : /
 Sub-directories : 
   /control/   UI control commands.
   /units/   Available units.
 Commands : 
PreInit> cd /control/
PreInit> ls
Command directory path : /control/

Guidance :
UI control commands.

 Sub-directories : 
   /control/cout/ Control cout/cerr for local thread
 Commands : 
   macroPath * Set macro search path ...
   execute * Execute a macro file.
   ...
\end{lstlisting}
where \verb|/gui/| does not exist, as it cannot be used in the TUI mode.

Note that in Geant4 the simulation exists in different states, which determine what commands are available and what the toolkit is currently doing. The prompt (e.g., \verb|PreInit>|) displayed in the terminal changes dynamically to reflect these states. \verb|PreInit| is the first state. Others will be introduced in a later section.

Users can also practice combining macro commands into a file, for instance, \emph{run.mac}:
\begin{lstlisting}[caption={A Geant4 macro file, \emph{run.mac}, specific for v0.}, label={l:macro}, language=G4macro]
# print important environment variables
/control/getEnv PATH
/control/getEnv GEANT4_DATA_DIR
/control/listAlias
\end{lstlisting}
This file can then be executed using the \verb|execute| command in the Geant4 interactive session:
\begin{lstlisting}
PreInit> /control/execute run.mac
\end{lstlisting}

For users who are more comfortable with the GUI, they can click the folder icon in the top-left corner of the Geant4 GUI window to load an auxiliary macro file, \emph{menu.mac}, to expand the menu bar of the GUI with frequently used commands using macro commands in the \verb|/gui/| directory. A user can simply select an item from the menu to execute a macro command.

The menu grows as the program evolves. At this stage, it only contains four top-level items: ``Execute'', ``Verbosity'', ``Window'', and ``Help''. In ``Help'', users can select ``Print Macro Commands on Screen'' or ``Generate Manual in HTML Format'' to view a list of  macro commands available at this stage and their descriptions. Curious users can open \emph{menu.mac} to see the macro commands corresponding to the menu items. The fully developed \emph{menu.mac} is shown in Listing \ref{l:menu}.

\begin{figure*}[htbp]\centering
\begin{lstlisting}[language=G4macro, basicstyle=\ttfamily\scriptsize, caption={Fully developed \emph{menu.mac} file for a customized Geant4 GUI menu bar.}, label=l:menu, numbers=left, frame=leftline, xleftmargin=5em]
# print each macro command on screen
/control/verbose

# clear existing menu
/gui/clearMenu

/gui/addMenu det Detector
/gui/addButton det "Check Volume Overlapping (After /run/initialize)" "/geometry/test/run"
/gui/addButton det "Create A Scorer for Volume" "/score/create/realWorldLogVol"
/gui/addButton det "Record Dose Deposit" "/score/quantity/doseDeposit"
/gui/addButton det "Record Energy Deposit" "/score/quantity/energyDeposit"
/gui/addButton det "Close Scorer" "/score/close"
/gui/addButton det "Dump Quality to File" "/score/dumpQuantityToFile"
/gui/addButton det "Draw Score Projection" "/score/drawProjection"
/gui/addButton det "Set Uniform B field" "/globalField/setValue"

/gui/addMenu exe Execute
/gui/addButton exe "A System Command" "/gui/system"
/gui/addButton exe "A Macro File" "/control/execute"

/gui/addMenu run Particle
/gui/addButton run "Set Particle" "/gps/particle"
/gui/addButton run "Set Energy" "/gps/energy"
/gui/addButton run "Set Direction" "/gps/direction"
/gui/addButton run "Set Position" "/gps/pos/centre"

/gui/addMenu run Run
/gui/addButton run "Initialize" "/run/initialize"
/gui/addButton run "Set Number of Threads" "/run/numberOfThreads"
/gui/addButton run "Set Number of Events" "/run/beamOn"
/gui/addButton run "10 Events" "/run/beamOn 10"
/gui/addButton run "100 Events" "/run/beamOn 100"
/gui/addButton run "1000 Events" "/run/beamOn 1000"

/gui/addMenu verb Verbosity
/gui/addButton verb "Disable EM Process Verbosity" "/process/em/verbose 0"
/gui/addButton verb "Disable Hadronic Process Verbosity" "/process/had/verbose 0"
/gui/addButton verb "Enable B-field Verbosity" "/globalField/verbose 1"
/gui/addButton verb "Enable Run Verbosity" "/run/verbose 2"
/gui/addButton verb "Set Macro Cmd Verbosity" "/control/verbose"
/gui/addButton verb "Set Tracking Verbosity" "/tracking/verbose"


/gui/addMenu vis Visualization
/gui/addButton vis "Open OpenGL Viewer" "/vis/open OGL"
/gui/addButton vis "Draw Detector Geometry" "/vis/drawVolume World"
/gui/addButton vis "Draw Axes" "/vis/scene/add/axes"
/gui/addButton vis "Draw GPS" "/vis/scene/add/gps"
/gui/addButton vis "Draw Scale" "/vis/scene/add/scale"
/gui/addButton vis "Draw B-field" "/vis/scene/add/magneticField"
/gui/addButton vis "Draw Trajectories" "/vis/scene/add/trajectories rich"
/gui/addButton vis "Set Background Color" "/vis/viewer/set/background"
/gui/addButton vis "Set Lights Vector" "/vis/viewer/set/lightsVector"
/gui/addButton vis "Set Up Vector" "/vis/viewer/set/upVector"
/gui/addButton vis "Set Viewpoint Vector" "/vis/viewer/set/viewpointVector"
/gui/addButton vis "Toggle Auxiliary Edges" "/vis/viewer/set/auxiliaryEdge"
/gui/addButton vis "Toggle Trajectory Accumulation" "/vis/scene/endOfEventAction"
/gui/addButton vis "Zoom in 20%" "/vis/viewer/zoom 1.2"
/gui/addButton vis "Zoom out 20%" "/vis/viewer/zoom 0.8"
/gui/addButton vis "Take Screenshot" "/vis/ogl/export detector.png"

/gui/addMenu window Window
/gui/addButton window Close exit

/gui/addMenu help Help
/gui/addButton help "Print Macro Commands on Screen" "/control/manual /"
/gui/addButton help "Generate Manual in HTML Format" "/control/createHTML"
/gui/addButton help "List Analysis Objects (After /run/initialize)" "/analysis/list 0"
/gui/addButton help "List Materials" "/material/nist/listMaterials all"
/gui/addButton help "List Particles" "/particle/list all"
/gui/addButton help "List Processes (After /run/initialize)" "/process/list all"
/gui/addButton help "List Units" "/units/list"
/gui/addButton help "List Scorers" "/score/list"
/gui/addButton help "List Scorer Color Maps" "/score/colorMap/listScoreColorMaps"
\end{lstlisting}
\end{figure*}

\subsection{Batch Execution}
The \verb|ui| branch introduced the interactive UI sessions. The \verb|batch| branch focuses on the fundamental C++ concept of distinguishing between interactive and non-interactive (batch) execution by checking command-line arguments.

A user can inspect the changes required to achieve this dual functionality by examining the difference in \emph{mingle.cc} between the two branches:
\begin{lstlisting}
git diff ui batch mingle.cc
\end{lstlisting}
The output highlights the modification to the \verb|main| function:
\begin{lstlisting}[language=G4C++, basicstyle=\ttfamily\scriptsize]
diff --git a/mingle.cc b/mingle.cc
...
+#include <G4UImanager.hh>
 int main(int argc,char** argv)
 {
-  G4UIExecutive ui(argc, argv);
-  ui.SessionStart();
+  if (argc == 1) { // start an interactive UI session
+    G4UIExecutive ui(argc, argv);
+    ui.SessionStart();
+  } else { // run a macro file in batch mode
+    G4String cmd="/control/execute ", macroFile=argv[1];
+    G4UImanager::GetUIpointer()
         ->ApplyCommand(cmd + macroFile);
+  }
 }
\end{lstlisting}
The number of arguments (\verb|argc|) is checked to determine the mode of execution. If there is only one argument, which is the program name, the program enters interactive mode, where it starts a Geant4 session waiting for user input. If there are two arguments, the program enters batch mode, where it executes a macro file specified by the second argument:

\begin{lstlisting}[language=Bash, deletekeywords={in}]
# program name itself is counted as one argument
mingle run.mac
ran in batch mode
\end{lstlisting}

The \verb|ui| version required the user to enter the interactive session to execute a macro file. In contrast, the \verb|batch| version does it using the \verb|G4UImanager| class \cite{G4UImanager} in C++. \emph{run.mac} shown below uses macro commands in \verb|/control/| directory to change the output message according to the execution mode:
\begin{lstlisting}[language=G4macro, basicstyle=\ttfamily\scriptsize]
/control/doifBatch /control/echo ran in batch mode
/control/doifInteractive /control/echo in interactive mode
\end{lstlisting}
where \verb|/control/echo| is used to print a message on screen.

\subsection{Run Management} \label{s:run}
The \verb|run| branch introduces a central component responsible for managing the simulation: the \emph{Run Manager}. The user can inspect the minimal C++ additions required to initialize the simulation kernel using the Git difference command:
\begin{lstlisting}
git diff batch run mingle.cc
\end{lstlisting}
The output highlights the modification to the \verb|main| function:
\begin{lstlisting}[language=G4C++, basicstyle=\ttfamily\scriptsize]
...
+#include <G4RunManagerFactory.hh>
... 
 int main(int argc, char** argv)
 {
+    auto run = G4RunManagerFactory::CreateRunManager();
     if (argc==1) { ... }
+    delete run;
 }
\end{lstlisting}

The code uses the modern \verb|G4RunManagerFactory| class \cite{G4RunManagerFactory} to create the manager. This approach teaches a critical modern Geant4 concept: the ability to switch between the default multi-threaded (MT) execution mode and the serial mode dynamically via the \verb|G4RUN_MANAGER_TYPE| environment variable. This allows the user to easily debug and compare MT performance. The following terminal outputs illustrate this seamless transition:
\begin{lstlisting}[caption={Default (Multi-threaded) Execution.}, label={l:mt}, basicstyle=\ttfamily\footnotesize]
mingle 
 Geant4 version Name: geant4-... [MT] (...)
 << in Multi-threaded mode >> 
 ...
\end{lstlisting}

\begin{lstlisting}[caption={Serial Execution (Overridden by Environment Variable).}, label={l:serial}, basicstyle=\ttfamily\footnotesize]
G4RUN_MANAGER_TYPE=Serial mingle
Environment variable "G4RUN_MANAGER_TYPE" 
enabled with value == Serial. 
Overriding G4RunManager type...

 Geant4 version Name: geant4-... [Serial] (...)
 ...
\end{lstlisting}

The addition of \verb|delete run;| at the end of the \verb|main| function teaches responsibility for proper C++ memory management, ensuring that resources allocated for the simulation kernel are correctly released upon program termination. This is a concept new to beginners who may be familiar with languages like Python, where memory management is handled automatically.

The instantiation of a run manager registers a comprehensive set of macro commands available to the user for the first time:
\begin{lstlisting}[basicstyle=\ttfamily\footnotesize, language=G4macro]
Command directory path : /
 Sub-directories : 
   /control/   UI control commands.
   /units/   Available units.
   /particle/   Particle control commands.
   /tracking/   TrackingManager and ...
   /geometry/   Geometry control commands.
   /process/   Process Table control commands.
   /event/   EventManager control commands.
   /cuts/   Commands for G4VUserPhysicsList.
   /run/   Run control commands.
   /random/   Random number status control commands.
\end{lstlisting}
This significantly expands the interactive control capabilities beyond the minimal \verb|/control/|,  \verb|/units/|, and \verb|/gui/| directories seen previously. However, many of them are merely space-holders that will only become useful when their corresponding components are implemented later.

A few macro commands do become useful at this stage, for example, \verb|/units/list| now can print out the list of units used by Geant4. It is added to \emph{menu.mac} as an item in the ``Help'' menu. More verbosity control commands are added to the ``Verbosity'' menu as well. \verb|/random/setSeeds| is added to \emph{run.mac} to demonstrate initializing the random number generator with an integer seed stream.

\subsection{Physics List}
A physics list defines what processes and how often certain particles undergo at which energy range. Utilizing the \verb|G4PhysListFactory| class \cite{G4PhysListFactory}, the \verb|physics| branch enables a beginner to select a physics list in the simplest, most recommended way. Only two lines of code are added on top of the \verb|run| branch to achieve this: line 38 and 52 in Listing \ref{l:mingle}.

\verb|G4PhysListFactory| allows beginners to obtain a meaningful and validated reference physics list without the low-level ``assembly'' of individual particles and processes. Users can switch the physics list at run-time using the \verb|PHYSLIST| environment variable, as demonstrated below:
\begin{lstlisting}[basicstyle=\ttfamily\footnotesize]
PHYSLIST=FTFP_BERT_EMV mingle
...
G4PhysListFactory::GetReferencePhysList
 <FTFP_BERT_EMV> EMoption= 1
<<< Geant4 Physics List simulation engine: FTFP_BERT

G4VModularPhysicsList::ReplacePhysics: G4EmStandard 
with type : 2 is replaced with G4EmStandard_opt1
\end{lstlisting}

With the physics list now defined, the application's macro commands are further expanded, enabling two new sets for the user: 1. \verb|/material/| for inspecting built-in materials, and 2. \verb|/physics_lists/| for toggling specific physics processes.

\verb|/particle/list| now can really be used to list all particles available in Geant4 and is added to \emph{menu.mac} together with \verb|/material/nist/listMaterials| as items in the ``Help'' menu. In addition, more sub commands become available in \verb|/process/| and \verb|/particle/| directories.

\subsection{Detector Definition}
The \verb|detector| branch introduces one of the mandatory user initialization components: the detector definition, through the definition of the \verb|Detector| class and its instantiation in line 53 of Listing \ref{l:mingle}. However, instead of the traditional approach of hard-coded volumes in C++, MinGLE utilizes the Geant4 text geometry format \cite{tg}, a standard feature of Geant4 since version 9.2, realized through the \verb|G4tgbVolumeMgr| class \cite{G4tgbVolumeMgr}.

As a simple example, a 10$\times$10$\times$10 $\text{m}^3$ cubic experimental hall filled with air can be defined in a text file named \emph{detector.tg} with just one line of text:
\begin{lstlisting}[language=Tg]
:volu hall BOX 5*m 5*m 5*m G4_AIR
\end{lstlisting}
In contrast, defining the same ``hall'' in C++ requires a multi-step process involving the instantiation of at least four different objects: solid (BOX), material (\verb|G4_AIR|), logical volume and physical volume (hall), often exceeding five lines of code. The \verb|:volu| syntax collapses this C++ hierarchy into a single, human-readable line, significantly reducing the "syntax noise" for those new to the toolkit.

Loading the detector definition from an external text file keeps the length and complexity of C++ code constant, regardless of the detector's complexity. Whether you are simulating a simple box or a complex detector with thousands of volumes, the C++ code remains a clean, 4-line interface (Line 8-11 in Listing \ref{l:mingle}).

Moreover, by separating the detector definition from the C++ source code, MinGLE becomes a universal engine. Users can modify dimensions or materials of the detector and re-run the simulation immediately. This skips the \emph{compile-link-run} cycle in a typical C++ program, allowing for rapid experimentation with how geometry affects physics results.

To clearly demonstrate all the benefits listed above, more than 100 lines of C++ code distributed in two files in Geant4 basic example B1 \cite{b1} are reduced to just 8 lines in \emph{detector.tg}:
\begin{lstlisting}[language=Tg, caption={Geant4 basic example B1 detector redefined in \emph{detector.tg}.}, label={l:detector}]
:volu World BOX 120 120 180 G4_AIR
:volu Envelope BOX 100 100 150 G4_WATER
:volu Shape1 CONS 0 20 0 40 30 0 360 G4_A-150_TISSUE
:volu Shape2 TRD 60 60 50 80 30 G4_BONE_COMPACT_ICRU

:rotm r000 0 0 0
:place Envelope 1 World  r000 0  0   0
:place Shape1 1 Envelope r000 0  20 -70
:place Shape2 2 Envelope r000 0 -10  70
\end{lstlisting}
which defines a cut cone (red) made of A-150 tissue-equivalent plastic and a trapezoid (green) made of compact bone, placed in a water-filled box (blue) as shown in Figure \ref{f:gui}.

To use the geometry defined in \emph{detector.tg}, the file must be placed in the directory where \verb|mingle| is executed. Otherwise, Geant4 will complain that the file does not exist.

Note that the external file name \emph{detector.tg} is hard-coded to keep the template concise and avoid implementing custom macro commands to load arbitrary filenames. For users who prefer to load files with other names, they can use the built-in \verb|#include| directive inside \emph{detector.tg} to point to another file in a different directory:
\begin{lstlisting}[language=Tg]
#include /path/to/my/favorite/experiment.tg
\end{lstlisting}

While GDML~\cite{gdml06} is another common standard for external geometry, it requires the Xerces-C library \cite{xerces-c}; text geometry is implemented instead of GDML in MinGLE because it provides the same decoupling benefits without adding extra dependencies, hence are more beginner-friendly.

With the detector defined, one can finally initialize a run using the following command without crashing the program:
\begin{lstlisting}[language=G4macro]
/run/initialize
\end{lstlisting}
which will load the detector geometry and initialize physics processes. After that, \verb|/geometry/test/run| can be used to check if there is any overlapping between detector volumes, and \verb|/process/list| can be used to list all initialized physics processes, as shown in \emph{run.mac} included below:

\begin{lstlisting}[caption={\emph{run.mac} in MinGLE v4.}, label={l:runv4}, language=G4macro, numbers=left, frame=leftline, xleftmargin=2em]
# print each macro command on screen
/control/verbose

# avoid intertwined outputs from threads
/control/cout/useBuffer

# disable massive output
/process/em/verbose 0
/process/had/verbose 0

# show in detail how a run is done
/run/verbose 2
# load detector definition and physics processes
/run/initialize

# check overlapping between detector volumes
/geometry/test/run

# print physics processes
/process/list
\end{lstlisting}

By default, \verb|/run/initialize| prints a lot of information on the screen, which may be overwhelming for a beginner. To avoid that, line 8 and 9 in \emph{run.mac} are used to disable the verbose output from electromagnetic and hadronic physics processes. On the other hand, line 12 is used to print in detail how a run is done on screen:

\begin{lstlisting}[language=G4macro, basicstyle=\ttfamily\scriptsize ]
...
# show in detail how a run is done
/run/verbose 2
# load detector definition and physics processes
/run/initialize
userDetector->Construct() start.
World is registered to the default region.
physicsList->Construct() start.
...
physicsList->CheckParticleList() start.
physicsList->setCut() start.
# check overlapping between detector volumes
/geometry/test/run
Running geometry overlaps check...
Checking overlaps for volume Envelope:1 (G4Box) ... OK!
Checking overlaps for volume Shape1:1 (G4Cons) ... OK!
Checking overlaps for volume Shape2:2 (G4Trd) ... OK!
Geometry overlaps check completed !
# print physics processes
/process/list
     Transportation,   GammaGeneralProc,   msc,   eIoni
     ...
G4 kernel has come to Quit state.
UserDetectorConstruction deleted 0xaa6e06de0
UserPhysicsList deleted 0x109c777d0
UserActionInitialization deleted 0x0
UserWorkerInitialization deleted 0x0
UserWorkerThreadInitialization deleted 0x0
UserRunAction deleted.
UserPrimaryGenerator deleted.
RunManager is deleting RunManagerKernel.
G4SDManager deleted.
EventManager deleted.
Units table cleared.
TransportationManager deleted.
Total navigation history collections cleaned: 2
G4RNGHelper object is deleted.
...
G4Allocator objects are deleted.
UImanager deleted.
StateManager deleted.
RunManagerKernel is deleted. Good bye :)
RunManager is deleted.
\end{lstlisting}
which shows clearly when each component is constructed and then deleted.

By default, Geant4 runs in multi-thread mode. Each thread prints its own output. To avoid intertwined outputs from different threads, line 5 in \emph{run.mac} is used to group the output from each thread into a buffer and print out one buffer at a time. However, this is still not ideal as a lot of information is duplicated in each thread buffer. The cleanest output can only be obtained by running \verb|mingle| in serial mode as demonstrated in Listing \ref{l:serial}.

\emph{menu.mac} is further expanded at this stage to allow users to execute individual commands in \emph{run.mac} using a Geant4 GUI.

A new macro command set \verb|/physics_engine/| appears after \verb|/run/initialize| is called. It allows users to set energy and time limits for neutron tracking.

In TUI, the change of Geant4 states is shown by different prompts before and after \verb|/run/initialize| is called:
\begin{lstlisting}
G4UI_USE_TCSH=1 mingle
...
PreInit> /run/initialize
...
Idle>
\end{lstlisting}
\verb|Idle| indicates that basic components, such as detector geometry, physics processes, etc., are loaded to the memory, and Geant4 is ready to run a simulation.

\subsection{Visualization}
The \verb|vis| branch enables the visualization of detector geometry at this stage and particle trajectories at later stages. This is achieved using the \verb|G4VisExecutive| class \cite{G4VisExecutive} as shown in line 41 and 59 in Listing \ref{l:mingle}.

The \verb|G4VisExecutive| is initialized with the ``quiet'' argument to suppress initial verbosity that may confuses a beginner. However, its verbose output can be printed on demand using \verb|/vis/list|, which is added to the ``Help'' menu in \emph{menu.mac}.

Part of the output of \verb|/vis/list| is a list of visualization drivers available to the user's specific Geant4 installation:
\begin{lstlisting}[caption={Common visualization drivers in Geant4.}, label={l:vislist}]
Registered graphics systems are:
  ASCIITree (ATree)
  DAWNFILE (DAWNFILE)
  RayTracer (RayTracer)
  VRML2FILE (VRML2FILE)
  ...
  TOOLSSG_OFFSCREEN (TSG_OFFSCREEN, TSG_FILE)
  OpenGLImmediateQt (OGLIQt, OGLI)
  ...
\end{lstlisting}
The OpenGL based visualization is demonstrated in Figure \ref{f:gui}.

ASCIITree does not really visualize anything. It prints the hierarchical structure of the detector geometry in a terminal, similar to that shown in the left panel of Figure \ref{f:gui}.

The usage of other visualization drivers in Listing \ref{l:vislist} is demonstrated in \emph{run.mac}. They all create an output file that is either simply an image file or a file that can be opened in a dedicated viewer.

A typical work flow is to first use the Geant4 GUI shown in Figure \ref{f:gui} to fine-tune the detector geometry visualization using mouse drags and zooms, and then use these other drivers to create an image file accordingly.

Crucially, the instantiation of \verb|G4UIExecutive| is refactored from a stack object to a pointer \cite{cpp}. This allows the visualization manager to be initialized \emph{after} the UI executive is created but \emph{before} the session starts. This specific initialization order is required since Geant4 version 11 to enable automatic visualization driver selection as demonstrated in the \emph{vis.mac} file in Geant4 basic example B1 \cite{b1}.

At this stage, \emph{menu.mac} is expanded with a new ``Visualization'' menu, providing convenient point-and-click access to common visualization commands as shown in the dropdown menu in Figure \ref{f:gui}.

The color and transparency of detector volumes can be adjusted through mouse clicks on the GUI, macro commands, or appended to the text geometry definition file \emph{detector.tg}:
\begin{lstlisting}[language=Tg]
:vis World OFF
:color Envelope 0.1 0.1 0.9 0.2
:color Shape1 0.9 0.1 0.1
:color Shape2 0.1 0.9 0.1
\end{lstlisting}
The first three numbers are the RGB color values. The last number is the transparency value. If it is not specified, the default value is 1.0 (fully opaque).

A new macro command set \verb|/param/| also appears after \verb|/run/initialize| is called, which is used for fast simulation (parameterization) control, a topic beyond the scope of MinGLE.

\subsection{Particle Generation}
The \verb|gps| branch implements the \verb|Generator| class (lines 20-28 in Listing \ref{l:mingle}), which utilizes the Geant4 component \verb|G4GeneralParticleSource| (GPS) \cite{G4GeneralParticleSource} to generate primary particles to start a simulation. It also implements the \verb|Action| class (lines 31-35 in Listing \ref{l:mingle}) to define the action of generating particles, which is used in the \verb|main| function (line 54 in Listing \ref{l:mingle}).

GPS allows users to define primary particles' properties, such as type, energy, position, and angular distribution, entirely through macro commands \cite{gps}, eliminating the need to hard-code specific particle guns in C++. Because it is far more flexible and powerful than the standard \verb|G4ParticleGun|, the \verb|G4GeneralParticleSource| class is utilized as the default particle generator in MinGLE.

In \emph{menu.mac}, a new ``Particle'' menu is added to simplify the configuration of common GPS settings through the GUI. A few more items are added to the ``Run'' menu to set the number of events of a simulation.

\emph{run.mac} in this version mimicks the Geant4 basic example B1 \cite{b1}, demonstrating how to shoot 210 MeV protons to the geometry and print simulation steps one by one on screen. Two PNG files will be generated in the current directory after running \verb|mingle run.mac|, one for the geometry and the other for geometry + particle trajectories, similar to the one shown in Figure \ref{f:gui}. 

\subsection{Histogram Generation}
The \verb|scorer| branch activates the built-in Geant4 component \verb|G4ScoringManager| \cite{G4ScoringManager} at line 47 of Listing \ref{l:mingle}. It allows users to define a scoring mesh (a 3D grid) over any volume in the detector to record physical quantities like energy deposition, dose, or flux without modifying C++ code.

\begin{figure}[htbp]\centering
\includegraphics[width=0.75\linewidth]{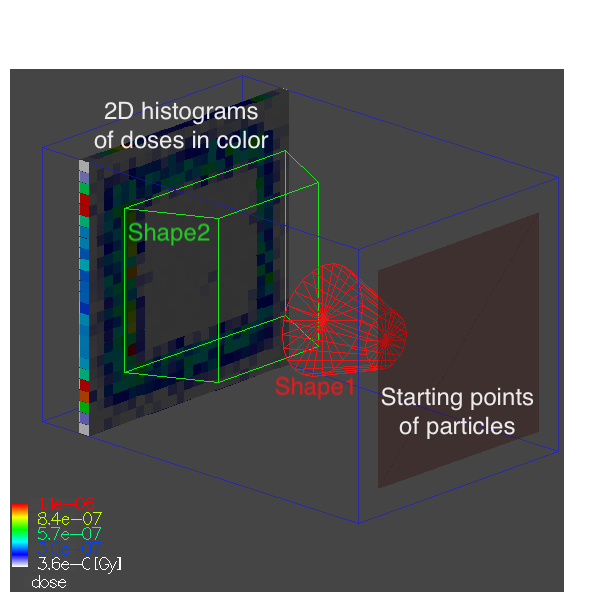}
\caption{Dose distribution behind \emph{Shape2} recorded by the scoring manager shown as 2D histograms in color.}\label{f:dose}
\end{figure}

Two types of meshes are demonstrated in \emph{run.mac} at this stage: one is identical to \emph{Shape1} defined in \emph{detector.tg}, it is used to record the number of protons going through it in a CSV file; the other is a $25\times25\times1$ grid used to record dose distributions right after \emph{Shape2} in a PNG file shown in Figure \ref{f:dose}. The gray region in the middle of the histogram facing \emph{Shape2} shows that many particles shot from the right red square toward \emph{Shape2} are absorbed by \emph{Shape2}.

In \emph{menu.mac}, the ``Detector'' menu is expanded to allow users to attach scorers to specific volumes via the GUI.

\subsection{Ntuple Generation}
While histograms are useful for quick checks, detailed analysis often requires event-by-event data. The \verb|ntuple| branch introduces the \verb|G4TScoreNtupleWriter| \cite{G4TScoreNtupleWriter} at line 48 of Listing \ref{l:mingle}. This class automatically saves data recorded by the scoring manager into an ntuple \cite{ntuple} in ROOT TTree \cite{ttree} format.

For multi-threaded runs, the writer is instructed to merge ntuples from different worker threads into a single output file, \emph{scoring.root} (lines 49-50 in Listing \ref{l:mingle}). This output can then be analyzed using the ROOT framework \cite{root} or Python libraries like Uproot \cite{uproot}.

The following \emph{run.mac} is used to demonstrate the recording of energy deposit in \emph{Shape1} event by event:
\begin{lstlisting}[language=G4macro, numbers=left, frame=leftline, xleftmargin=2em]
/run/initialize

# use 60 keV gamma to check energy spectrum
/gps/particle gamma
/gps/energy 60 keV
/gps/direction 0 0 1
/gps/pos/centre 0 2 -11 cm

# record energy deposit in Shape1
/score/create/realWorldLogVol Shape1 2
/score/quantity/energyDeposit e
/score/close

/run/printProgress 1000
/run/beamOn 10000
\end{lstlisting}
The full absorption peak of 60 keV gamma rays instead of high energy protons are used to check whether the energy deposit is recorded correctly. The last number in line 10 is the depth of \emph{Shape1} in the detector hierarchy instead of its copy number.

The name of the ntuple in general is \verb|mesh_quantity|, where \verb|mesh| is the name of the scoring mesh, in this case, \emph{Shape1}, and \verb|quantity| is the name of the quantity to be recorded, in this case, \verb|e| for energy deposit. And the name of the branch in the ntuple that records the event by event energy deposit is \verb|Shape1_e_score|. All of these can be checked by the following ROOT commands (\verb|.ls| and \verb|TTree::Show()|):
\begin{lstlisting}[language=Root]
root scoring.root
root[0] .ls
TFile**		scoring.root	
 TFile*		scoring.root
  KEY: TTree	Shape1_e;1	Shape1_e
root[1] Shape1_e->Show(0)
======> EVENT:0
 Shape1_e_eventId = 1
 Shape1_e_cell   = 0
 Shape1_e_score  = 0.0239774
root[3] Shape1_e->Draw("Shape1_e_score")
\end{lstlisting}
The last command will fill a histogram with the energy deposit in \emph{Shape1} event by event, demonstrating the effectivenss of using the ROOT framework to analyze the ntuple.

\begin{figure}[htbp]\centering
\includegraphics[width=\linewidth]{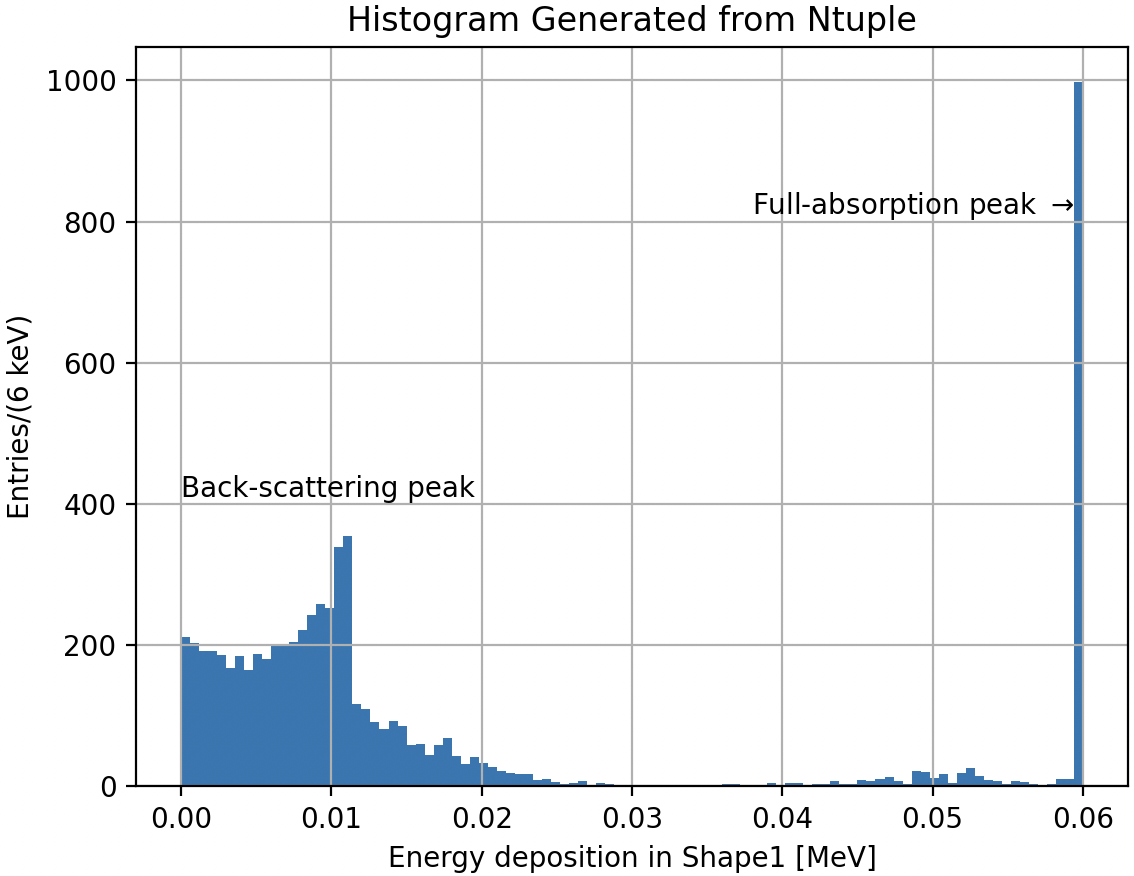}
\caption{Energy deposit in \emph{Shape1} from 60 keV gamma rays.}\label{f:e}
\end{figure}

A Python script \emph{drawE.py} is provided to draw the same distribution using uproot \cite{uproot} and matplotlib \cite{matplotlib} for users who are not familiar with ROOT. The result is shown in Figure \ref{f:e}.

Because the \verb|G4TScoreNtupleWriter| class inherits from \verb|G4AnalysisManager| \cite{G4AnalysisManager}, after \verb|/run/initialize|, a new macro command directory \verb|/analysis/| appears in the Geant4 UI.

The \verb|G4TScoreHistFiller| class \cite{G4TScoreHistFiller} also inherits from the \verb|G4AnalysisManager| class. It can be used to fill histograms during a run, but \verb|G4TScoreNtupleWriter| is chosen over \verb|G4TScoreHistFiller| since an ntuple can be used to create a histogram with a custom binning after the run.

\emph{menu.mac} is extended to allow users to pick a volume using the Geant4 GUI and record the energy deposit in it.

\subsection{Magnetic Field Setup}
The final stage, the \verb|field| branch, adds the ability to set up a uniform magnetic field throughout the detector.

\begin{figure}[htbp]\centering
\includegraphics[width=0.6\linewidth]{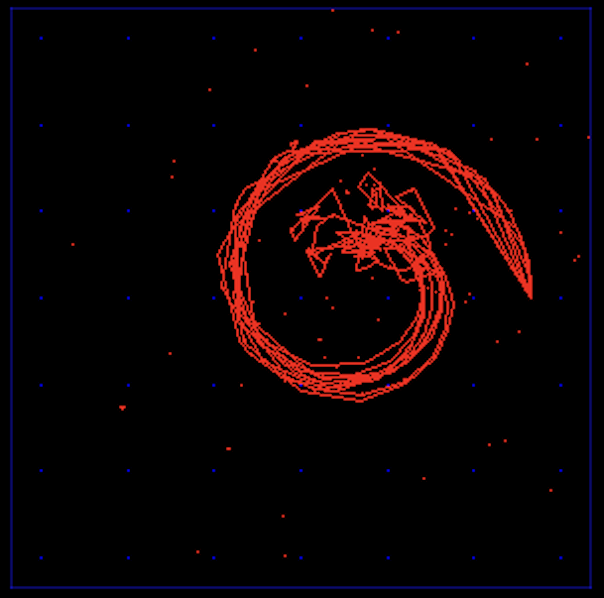}
\caption{Charged particle trajectories bending in a uniform magnetic field.}\label{f:bending}
\end{figure}

\verb|G4GlobalMagFieldMessenger| \cite{G4GlobalMagFieldMessenger}, registered in line 12-14 of Listing \ref{l:mingle}, provides a set of macro commands under the directory \verb|/globalField/|, allowing users to define a uniform magnetic field vector:
\begin{lstlisting}[language=G4macro]
/globalField/setValue 0 0 1 tesla
\end{lstlisting}

\emph{run.mac} is fine-tuned to show the bending of 100 MeV $\mu^-$ trajectories (red) in the B-field throughout the water filled box (black) as shown in Figure \ref{f:bending}.

\subsection{The \texttt{main} Branch}
The \texttt{main} branch serves as the final, production-ready state of the repository. While the core application code, \emph{mingle.cc} and \emph{menu.mac}, is identical to that in the \texttt{field} branch (representing the final stage of the incremental development), auxiliary files such as \emph{run.mac} and \emph{drawE.py} are different, as they are tailored for a more general demonstration rather than focusing on the introduction of a specific feature.

The distribution of files across branches in the MinGLE repository follows a structured strategy:
\begin{itemize}
  \item \emph{CMakeLists.txt} and \emph{.gitignore} are identical in all branches to provide a consistent build experience.
  \item \emph{mingle.cc}, \emph{detector.tg}, and \emph{menu.mac} grow gradually, with each development stage adding only the necessary lines to implement a new feature.
  \item \emph{README.md} (the ``instruction'') and \emph{run.mac} (the ``exercise'') are branch-specific to provide localized documentation and testing commands for each milestone.
  \item Certain project-level files, such as the MinGLE logo (\emph{logo.png}) and the project \emph{LICENSE}, exist only in the \texttt{main} branch to keep the learning-focused branches as minimalist as possible.
\end{itemize}

\section{Discussion}
The official Geant4 User's Guide includes a ``Getting Started'' section \cite{g4start} that introduces novice developers to essential Geant4 components through a series of C++ code snippets. These snippets are invaluable for understanding the specific syntax and purpose of individual classes. However, they are often presented in isolation, leaving the beginner without a cohesive, working program to immediately compile and experiment with.

MinGLE effectively fills this gap. Rather than providing static snippets to be manually assembled, MinGLE offers a complete, functional simulation kernel from the very first stage (interactive UI). As the user progresses through the roadmap, they are not just reading about classes like \verb|G4GeneralParticleSource| or \verb|G4tgbVolumeMgr|; they are observing them operational within a verified, compilability-tested environment. This transformational approach shifts the learning experience from passive code-reading to active application-building, providing the crucial ``missing link'' between the official documentation's building blocks and a finished toolkit implementation.

\section{Conclusion}
MinGLE offers a modern, minimalist paradigm for Geant4 application development. By stripping away boilerplate and utilizing contemporary toolkit features, it reduces the functional core of a simulation to its absolute essentials without sacrificing flexibility or power. The unique Git-based pedagogical structure addresses the ``missing link'' in official documentation by providing a continuous, compilable path from a simple UI to a feature-rich application. This approach ensures that novice developers can master the toolkit's complex assembly through active building rather than passive observation. Distributed as an open-source template under the MIT license, MinGLE is intended to serve as the definitive, extensible foundation for both educational purposes and new scientific research projects in the Geant4 ecosystem.

\section{Acknowledgements}
This work is supported by the NSF award OIA-2437416, PHY-2411825, and the Office of Research at the University of South Dakota. Computations supporting this project were performed on High Performance Computing systems at the University of South Dakota, funded by NSF award OAC-1626516.

\section{Declaration}

During the preparation of this work the author used \texttt{Gemini} to generate the initial draft of this manuscript, polish the language, and improve the formatting. After using this tool, the author reviewed and edited the content as needed and takes full responsibility for the content of the published article.

\bibliographystyle{elsarticle-num}
\bibliography{ref}
\end{document}